\documentstyle[preprint,prl,aps]{revtex}
\input {epsf.tex}
\begin{document}
\draft
\title{Rods Near Curved Surfaces and in Curved Boxes}
\author{K. Yaman, M. Jeng and P. Pincus\\}
\address{Department of Physics,
University of California Santa Barbara, CA 93106--9530}
\author{C. Jeppesen\\}
\address{ Materials Research Laboratory, 
University of California, Santa Barbara, CA 93106}
\author{ C. M. Marques\\}
\address{R.P.-C.N.R.S., Complex Fluid Laboratory UMR166, Cranbury,
NJ 08512--7500, USA}
\date{\today}
\maketitle
\begin{abstract}
We consider an ideal gas of infinitely rigid rods
near a perfectly repulsive wall, and
show that the interfacial tension of a surface with
rods on one side is lower when the surface bends towards the rods.
Surprisingly we find that rods on both sides of
surfaces also lower the energy  when the surface bends.
We compute the
partition functions of rods confined to spherical and
cylindrical open shells, and conclude that spherical
shells repel rods, whereas cylindrical shells (for thickness of the
shell on the order of the rod-length) attract them.
The role of flexibility is investigated
by considering chains composed of two rigid
segments.
\end{abstract}
\pacs{87.22.Bt, 11.30.Qc, 82.70.Dd}

\section{Introduction}
\label{sec:introduction}

\narrowtext

Macromolecular surface interactions play an important role
in a variety of phenomena:
stabilization/destabilization of colloidal solutions \cite{russel},
where the ``surface'' is the surface of a colloidal particle;
surface effects in liquid crystal displays \cite{degennes2},
where the ``surface'' is the walls of the container;
and modification of bending rigidities of bilayers \cite{safran},
in which case the ``surface'' is a membrane.

Both rigid and flexible macromolecules can induce a modification of
the properties of interfaces, either by adsorption or depletion;
here we concentrate on the surface physics of non-adsorbing,
rigid and  rod-shaped macromolecules.
Interest in solutions of rigid rod-like objects can be traced to the
work of Stanley who first
extracted the tobacco mosaic virus (TMV)~\cite{stanley} and to the
subsequent observation of the nematic phase in TMV
solutions~\cite{bawden}, which was
later explained by the seminal theory of
Onsager~\cite{onsager}. Many other rod-like molecules exist in
the biological realm, ranging from DNA in its
$\alpha$-helix form \cite{livolant} to fibrils of amyloid
$\beta$-protein, the molecular agent at the origin of the Alzheimer
disease \cite{lomakin}.
Solutions of rod-like molecules in a large variety of mineral and organic
systems have also been studied~\cite{buining}.

In a pioneering study of {\em surface} interactions
of macromolecules
Asakura and Oosawa~\cite{asakura} showed that the steric depletion
of rigid macromolecules
at a flat surface increases the interfacial energy of
the system, implying that two surfaces in a solution of
non-interacting rods will attract
one another when they are at separations
smaller than the rod-length. This effect, which is at the origin of
colloidal destabilization,  is also present in the steric
depletion of spherical particles, or particles with
other shapes \--- but with a different range and
magnitude.
Asakura and Oosawa theory has since been
extended  to include effects of nematic ordering
in the bulk~\cite{poniewierski}
or effects of  rod-rod excluded volume interactions~\cite{mao}; in the
latter work the authors show that when the
modification of the rod-rod excluded volume
interactions close to the surface is taken into
account, a rod solution may {\em stabilize} a colloidal suspension,
rather than destabilizing it.
However, almost all the available studies on rod-surface interactions
are devoted to flat geometries,
except for Auvray's work~\cite{auvray} where the author makes
a first attempt at
considering curved geometries; we use some of his constructions in
our work, which extends the original Asakura and Oosawa theory
to include effects of surface curvature in the contribution
of the rod depletion layer to the surface tension. For flexible
surfaces, e.g. membranes, in rod solutions,
our results allow us to
derive the renormalization of the bending rigidities of the membrane.
Some of our findings are qualitatively different from the corresponding
results for flexible polymers in solution.
This prompted us to consider two
rigid rod-segments joined in the middle as a
first step in studying the cross-over from rigid rods to
flexible polymers \---
rigid rods being a polymer of which the
persistence length is longer than its length, and the
rod-segments joined in the middle being one with
a persistence length equal to half of its length.
Not surprisingly we find that
when one allows for a hinge in the rod,
the results are closer to the flexible polymer results.

Even though steric effects are purely entropic and exist for any
non-adsorbing
macromolecular solution, the strength of the effects is larger for
rods with large aspect ratios.
The following simple comparison of the order of magnitudes of the
contributions to the surface tension with those caused by spherical
particles illustrates this.
The typical scale of the energy density
in a solution of  colloidal spheres of radius $r_0$ and
particle number density $\rho_b$ is
$k_B T \rho_b$. Integration over the layer thickness results in
corrections to the interfacial energy of the order of
$\Delta \gamma \simeq k_B T \rho_b r_0$. On the
other hand, the scale for the
interfacial tension in most liquids is of the order of
$\gamma_0\simeq k_B T/a^2$ where $a$ is a microscopic size. For
instance for $a\sim 0.1$nm, $\gamma_0$  is of  the order of tens of
mN/m.
The depletion of spherical particles,
even at order unity volume fractions $\phi = \rho_b (4 \pi/3) r_0^3$,
induces thus corrections to the interfacial tension which are
a factor $(a/r_0)^2$ lower than typical values.
Curvature corrections
to the interfacial energy are of the form:
$\Delta \gamma \simeq k_B T \rho_b r_0 ( 1 + C_1 r_0/R +
C_2 r_0^2/R^2) $, where $R$ is
the radius of curvature and $C_1, C_2$ are two
numerical constants. This sets the order of magnitude of
modifications of the bare  elastic
constants: $\Delta \kappa \simeq k_B T \rho_b r_0^3$. Even at the
upper concentration limit $\rho_b\sim 1/r_0^3$, these corrections are
at the lower end of the
range\cite{nelson}, $1-20 k_B T$, of most bare elastic
constants. For an isotropic rod solution
the upper concentration limit is
the Onsager concentration $\rho_b^\star=4.2/(L^2 D)$ \cite{onsager},
where
$L$ is the length of the rod and $D$ its diameter. Now interfacial
tension contributions are of the order $\Delta \gamma \simeq k_B
T \rho_b L$, but even for rod diameters of order of the microscopic
length $a$, and at the Onsager concentration
this contribution is still a factor ($a/L$)
smaller than typical interfacial tension values. However,
modifications of the elastic constants are here of order
$\Delta \kappa \simeq k_B T \rho_b L^3$. At the Onsager concentration
this is a factor $(L/D)$ larger than $k_B T$, and
therefore, may lead to a substantial modification of the elastic
constants of even rather rigid phospholipid membranes with elastic
constants as large as $20\,k_B T$, even
at low rod concentrations where rod-rod interactions are
negligible.

In the next section we introduce
the basic thermodynamics necessary for
our discussion; sections~\ref{sec:onesurface}
and~\ref{sec:twosurface} contain our results
for rods close to surfaces, and rods in shells,
respectively. We make an
initial attempt at understanding the cross-over from flexible polymers
to rigid rods in section~\ref{sec:segmented};
the last section is devoted to a
discussion and summary of our main results.

\section{Thermodynamics}
\label{sec:thermodynamics}

We consider an ideal gas of rods of length $L$ in the presence of
surfaces which repel the rods, in several different geometries. The
thickness of the rods is taken to be zero, i.e. much smaller than all
other length scales in the problem.
We parametrize the possible rod configurations by the coordinate
${\vec{r}}$, and  two angles specifying in which direction  the rod points,
${\vec{\omega}} \equiv {\vec{\omega}} (\theta, \phi)$.  Kinetics is ignored
throughout the  treatment\footnote{Kinetic effects will be smaller by a factor
of $(\rm{thermal \, wavelength} / L) \sim 0$.}.

The relevant potential describing the
thermodynamics of the system can be written as, ($k_B T \equiv 1$):
\begin{equation}\label{om}
\Omega\left[\rho({\vec{r}}, {\vec{\omega}})\right] = \int d\,{\vec{r}} \int
d\, {\vec{\omega}} \,\rho({\vec{r}}, {\vec{\omega}}) \left[\log(v \,
\rho({\vec{r}},
{\vec{\omega}})/e) - (\mu_b - U_{ext} ({\vec{r}},
{\vec{\omega}})) \right]
\end{equation} 
where $v$ is some normalization volume, $\mu_b$ is the solution
chemical potential, and  $U_{ext}$ is the hard wall interaction
potential that is either infinite or zero, depending on whether the
configuration of the rod is allowed by the `hard wall' requirement or
not. Functional minimization of $\Omega$ with respect to
$\rho({\vec{r}},{\vec{\omega}})$ gives the equilibrium density profile:
\begin{equation}  
\rho({\vec{r}}, {\vec{\omega}})  = {e^\mu \over v} e^{-U_{ext} ({\vec{r}},
{\vec{\omega}})} \equiv \frac{\rho_b}{4 \pi} e^{-U_{ext} ({\vec{r}},
{\vec{\omega}})}
\label{rhobar}
\end{equation}  
with $\rho_b$ the rod number density of the bulk solution. The local, position
dependent, number density of particles
$\rho({\vec{r}})$ is simply the sum over all allowed angular
configurations of $\rho({\vec{r}}, {\vec{\omega}})$.
The excess surface energy is then calculated as follows:
\begin{equation}
\label{delta} 
\Delta \gamma ={\Omega\left[\rho({\vec{r}}, {\vec{\omega}})\right]
-\Omega\left[\rho_b/(4 \pi)\right]\over S} =
\rho_b  \int {d\vec{r}\over{S}}\;\int {d\vec{\omega}\over{4\pi}}(1-
e^{-U_{ext} ({\vec{r}}, {\vec{\omega}})})
\label{delgam}
\end{equation} 
where $S$ is the surface area and the volume integral runs over the space
available to the solution.

\section{One Surface in a Rod Solution}
\label{sec:onesurface}

Here we consider surfaces with rods on one side only. Results for the case
with rods on both sides can be easily found by adding  the results for the
inside and the outside. Our calculations ignore fringe effects that are
unimportant in the thermodynamic limit, e.g. the flat wall is treated as
translationally invariant.

\begin{figure} [t]
\epsfxsize=5.0in
\epsfysize=6.0in
\vskip 2truecm
\centerline{\epsfbox{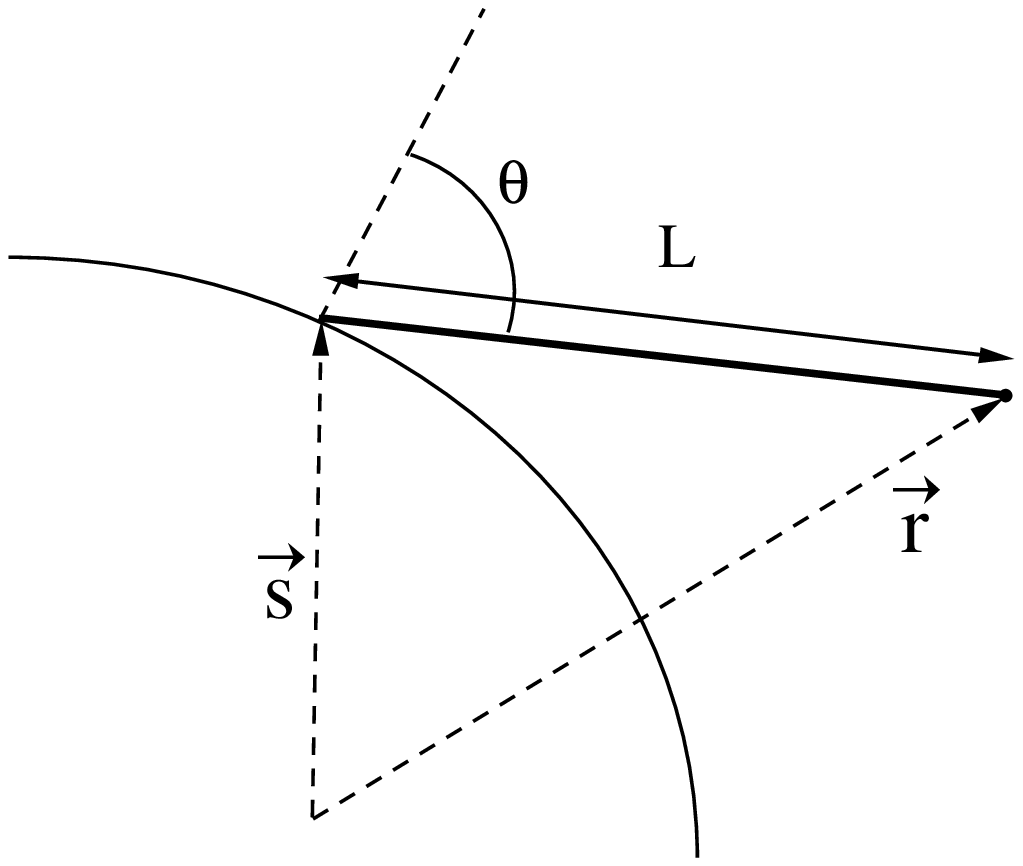}}
\vfil
\vskip -2truecm
\caption{Configurational space of  a rod near a  surface.
The rod is represented at the angle of contact. }
\label{figendcoord}
\end{figure}

We first rewrite equation (\ref{delgam}) in pressure units ($k_B T \rho_b =1$)
and explicitly state its rod-length dependence
\begin{equation}
\Delta \gamma (L)= \int {d\vec{r}\over{S}}\;\int
{d\vec{\omega}\over{4\pi}} \, h({\vec{r}}, {\vec{\omega}},L)
\label{delgamgen1}
\end{equation}
If we represent the space coordinates of the rod by the position of
one extremity (see figure~(\ref{figendcoord})), the function
$h({\vec{r}}, {\vec{\omega}},L)$ is unity when a rod  of length $L$,
starting at $\vec{r}$ and pointing in the   direction ${\vec{\omega}}$
intersects the surface; it is 0 otherwise. 
This can be written as $h({\vec{r}},
{\vec{\omega}},L)= H(L - (\vec{s}-\vec{r}) \cdot \vec{\omega})$, 
with $H(x)$ the
usual step function $H(x>0) = 1\, ; \, H(x<0) =0$ and $\vec{s}$ is
defined as follows:
The half-ray orginating at $\vec{r}$ and travelling in the direction
of $\vec{\omega}$ may or may not intersect the surface. If it does not
then we do not have to define $\vec{s}$, since all such configurations
give $H = 0$, and thus do not affect any of the below expressions. 
If the half-ray does intersect the surface, we define $\vec{s}$ as the
intersection point that is closest to $\vec{r}$. Notice that $\vec{s}$
is independent of $L$.

It is convenient to calculate the $L$-derivative of
the excess surface energy:
\begin{equation}
{\partial\over \partial L} \Delta \gamma (L)= \int {d\vec{r}\over{S}}\;\int
{d\vec{\omega}\over{4\pi}} \, 
\delta (L - (\vec{s}-\vec{r})\cdot\vec{\omega}) =
\int {d\vec{s}\over{S}}\;\int {d\vec{\omega}\over{4\pi}} \,
\cos\theta_{\vec{s}} \, g({\vec{s}}, {\vec{\omega}},L)
\label{delgamgen2}
\end{equation}
with $\theta_{\vec{s}}$ the angle between the rod and the normal to the surface
at point $\vec{s}$ and $ g({\vec{s}}, {\vec{\omega}},L)$ a function
restraining the angular integration space for a rod with one extremity at
$\vec{s}$. The integral over $d\vec{s}$ ranges over the whole surface.
The factor $\cos\theta_{\vec{s}}$ is the appropriate Jacobian.
 
\subsection{Convex bodies}
\label{sec:convex bodies}

For any surface delimiting a convex body, the function $ g({\vec{s}},
{\vec{\omega}},L)$  restricts the  angular  integral to half of the angular
space, independently of the actual surface shape.
\begin{equation}
{\partial \over \partial L} \Delta \gamma (L)= {1\over{4\pi}}
\int_0^{2\pi} d\phi
\int_0^{\pi/2}
d\theta \sin \theta  \,   \cos \theta\, = {1 \over 4}
\label{delgamcon}
\end{equation}
Integrating equation (\ref{delgamcon}) with the boundary
condition $\Delta
\gamma (L=0) = 0$ one gets  the three dimensional result for the surface
tension of convex bodies immersed in a rod solution:
\begin{equation}
\Delta \gamma (L)=  {L\over 4}
\label{delgamcon3}
\end{equation}
This result can easily  be extended to convex bodies in an arbitrary dimension
${\cal D}$:
\begin{equation}
\Delta\gamma(L) = c({\cal D})\, L
\label{linear3}
\end{equation}
with
\begin{equation}
c({\cal D}) = \frac{\Gamma({\cal D}/2)}
	{2 \sqrt{\pi}\, \Gamma(({\cal D} + 1)/2)} =
	\frac{({\cal D} - 2)!!}{({\cal D} - 1)!!} \times \left\{
\begin{array}{llll}
\frac{1}{2}   & \mbox{; for} & {\cal D} & \mbox{odd}\\
\frac{1}{\pi} & \mbox{; for} & {\cal D} & \mbox{even}
\end{array}
\right.
\label{constant}
\end{equation}
In particular, $c(2)=1/\pi$ and $c(3)=1/4$.

\subsection{General Surfaces}
\label{sec:general surfaces}

\begin{figure} [t]
\epsfxsize=5.0in
\epsfysize=4.0in
\centerline{\epsfbox{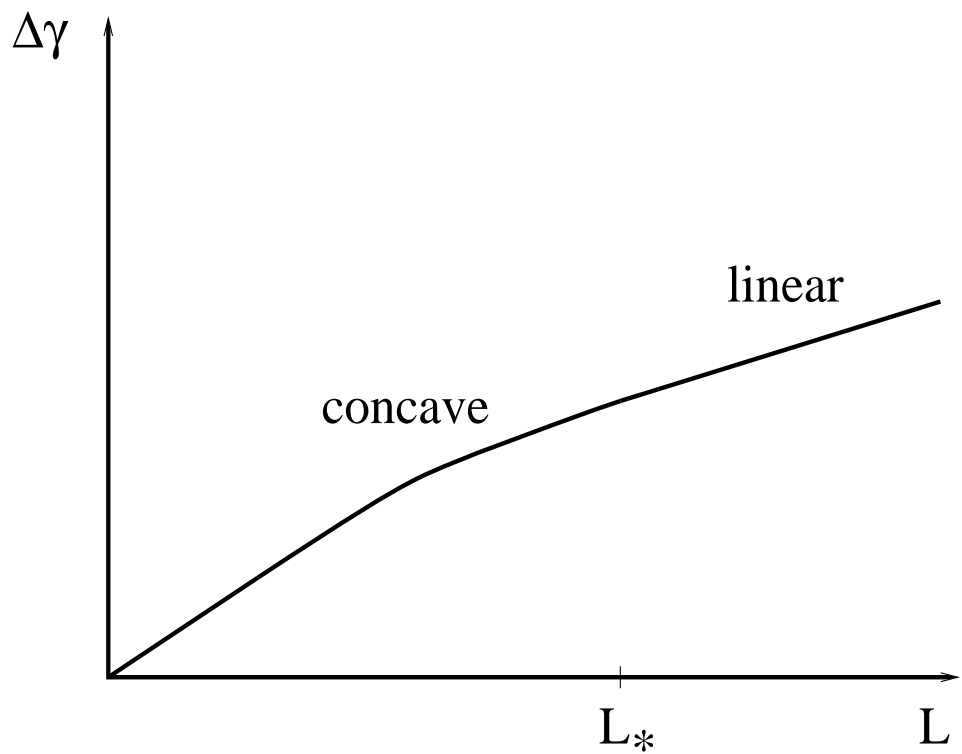}}
\vfil
\caption{General shape of the surface tension
as a function of the rod-length; $L_*$ is the maximum length of a
line that connects any two points on the surface, without intersecting
the surface.}
\label{figgamgen}
\end{figure}

For a general surface, the function $ g({\vec{s}},{\vec{\omega}},L)$
might restrict  the angular integration in equation (\ref{delgamgen2}) to less
than half of the angular space. In this case we can still integrate over half
of the angular space and further subtract the non-available angular sector.
In three dimensions:
\begin{equation}
{\partial \over \partial L} \Delta \gamma (L)= {1\over 4} -  \int
{d\vec{s}\over{S}}\;\int {d\vec{\omega}\over{4\pi}} \,
\cos\theta_{\vec{s}}
\, H(L-({\vec{s'}}-{\vec{s}}).{\vec{\omega}})
\label{ddelgaml}
\end{equation}
where ${\vec{s'}}$ is another surface vector. 
Differentiating with respect to
$L$ and performing the angular integration leads to
\begin{equation}
{\partial^2 \over \partial L^2} \Delta \gamma (L)= -  \int
{d\vec{s}\over{S}}\;\int {d\vec{\ell}\over 4\pi L^2} \,
\cos \theta_{\vec{s}} \cot \theta_{\vec{\ell}}
\label{ddelgamlres}
\end{equation}
The vector $\vec{\ell}$ defines all the points on the surface  at a distance
$L$ from a given point ${\vec{s}}$. In this three dimensional case this is
usually a one dimensional curve. The angle  $\theta_{\vec{s}}$ is measured
with respect to the normal of the surface at point ${\vec{s}}$ and the angle
$\theta_{\vec{\ell}}$ is measured with respect to the surface normal at point
$\vec{\ell}$. Contrary to the convex case, the excess surface energy of a
non-convex body immersed in a rod solution does depend on the actual
shape of the body. Moreover, because the contribution of the second order
derivative (\ref{ddelgamlres}) is always negative, any long-wavelength
concavity  will {\em reduce} the surface tension of  a surface with rods on
one side or both sides. Therefore entropic effects in such
systems will render flat surfaces unstable towards long-wavelength
curvature! It also implies that for a {\em fixed geometry} the surface
tension is a concave function of the rod-length, $L$, and will have
the general form depicted in figure~(\ref{figgamgen}).

Equation (\ref{ddelgamlres}) can easily be extended to spaces of different
dimensions. For instance, in two dimensions, where the ``surface'' is actually
a one dimensional curve of perimeter $P$, one has
\begin{equation}
{\partial^2 \over \partial L^2} \Delta \gamma (L)= -  \int
{d\vec{\ell}\over{P}}\;\sum_{\vec{q}} {1\over 2\pi L} \,
\cos \theta_{\vec{\ell}} \cot \theta_{\vec{q}}
\label{ddelgamlrestwo}
\end{equation}
The sum runs over all (usually two) points  ${\vec{q}}$ at a distance $L$ from
a given point ${\vec{\ell}}$ on the surface.

\subsection{Some examples}
\label{sec:examples}

With the above methods,
$\Delta\gamma(L)$ can be calculated exactly for the inside
and outside of a circle, a sphere, and a cylinder,
each with radius $R$; and also for two
flat walls separated by $\delta$,
i.e. a gap. We would like to emphasize that
{\em all} of these results can also be derived in an
independent way which is also exact, namely by ``counting beans'',
i.e. by just doing the integrals in Eq.~(\ref{delgamgen1})
straightforwardly (see section \ref{sec:comments} below). We have explicitly
checked that all of the  results quoted here agree with these straightforward
calculations \--- especially in the cylinder's case both calculations are
quite  involved algebraically, and this check is absolutely non-trivial. Of
course, the method of computing the second derivative and then integrating up
is  the shortest route to these answers.

We now define $\epsilon \equiv L/R$; this is for convenience only and does not
mean that we take $\epsilon$ to be small unless specifically stated.

\begin{figure} [t]
\epsfxsize=4.0in
\epsfysize=4.0in
\centerline{\epsfbox{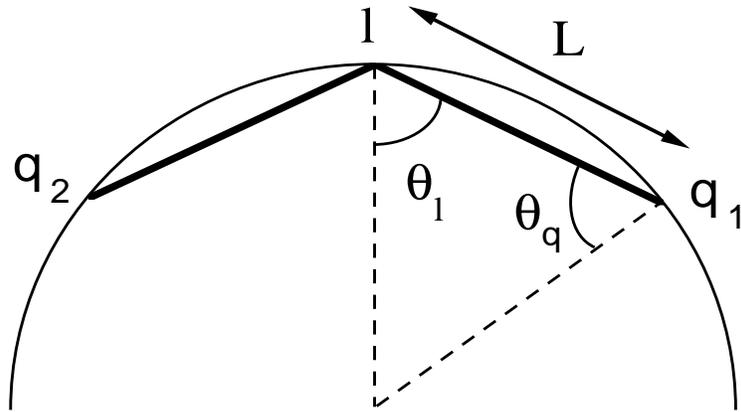}}
\vfil
\caption{Rods inside a circle.}
\label{figcircle}
\end{figure}

We first calculate results for two dimensions. Inside of a circle we have,
(see figure~(\ref{figcircle})):
$ \cos\theta_{\ell} = \epsilon/2$, $\cot\theta_q = \cot\theta_\ell=
(\epsilon/2)/\left(\sqrt{1 - \epsilon^2/4}\right)$,
and the sum over $q$ amounts to a factor of two. Thus:
\begin{equation}
{{\partial^2\, \over \partial L^2} \Delta\gamma^{in}_{cir}} =
-\frac{1}{\pi\,\, L}\,
\cos\theta_{\ell}\, \cot\theta_q = -\frac{\pi}{R}\,
\frac{\epsilon/4}
{\sqrt{1 - \epsilon^2/4}}
\label{secondderivativecircle}
\end{equation}
Using $\Delta\gamma(L=0) = 0$, 
and ${\partial \over \partial L}\,\Delta\gamma\, (L=0) = \frac{1}{\pi}$,
we find:
\begin{equation}
{\Delta \gamma^{in}_{cir} \over L} =
{1\over{\pi\,\epsilon}}\sin^{-1}\left(\epsilon/2\right) +
 {1\over{2\pi}}\sqrt{1-\epsilon^2/4}
\label{circle}
\end{equation}

\begin{figure} [t]
\epsfxsize=4.0in
\epsfysize=4.0in
\centerline{\epsfbox{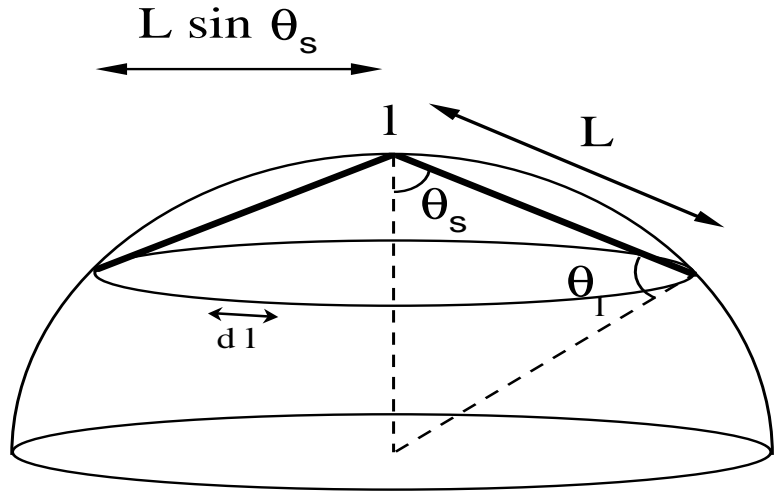}}
\vfil
\caption{Rods inside a sphere.}
\label{figsphere}
\end{figure}

For three dimensions we consider rods inside spheres and cylinders. 
Inside of a
sphere we have, (see figure~(\ref{figsphere})):
$\cos\theta_s = \epsilon/2$, $\cot\theta_\ell = \cot\theta_s =
(\epsilon/2)/\left(\sqrt{1 - \epsilon^2/4}\right)$ 
and $\ell = 2 \pi L\, \sin\theta_s$.
Thus:
\begin{equation}
{{\partial^2\, \over \partial L^2} \Delta\gamma^{in}_{sph}} =
-\frac{1}{2 L}\,
\sin\theta_s\, \cos\theta_s\, \cot\theta_s  =
- \frac{1}{8}\frac{\epsilon}{R}
\label{secondderivativesphere}
\end{equation}
It follows that:
\begin{equation}
\frac{\Delta \gamma^{in}_{sph}}{L} =
{1\over 4}-{1\over{48}}\,\epsilon^2
\label{sphere}
\end{equation}
The calculation for the inside of a cylinder is a bit tedious, but
also doable. We find:
\begin{equation}
{{\partial^2\, \over \partial L^2} \Delta \gamma^{in}_{cyl}} =
\frac{8}{3\pi\, R\,  \epsilon^3}\,
   \left[ \left(1+{{\epsilon^2}\over 4}\right)
	E\left[{{\epsilon^2}\over 4}\right] -
      \left(1+{{\epsilon^2}\over 8}\right)
	K\left[{{\epsilon^2}\over 4}\right] \right]
\label{cylinder}
\end{equation}
where the functions $E$ and $K$ are the complete elliptic integral of
the first and second kind respectively.
The cylinder result  can  in principle be integrated to give the special
functions MeijerG. The straight-forward ``bean-counting''
mentioned above leads also to an integral form that can only be done
explicitly as a perturbation series in $\epsilon$.  We checked that
the coefficients in this series agree with the expression for the second
derivative in Eq.~(\ref{cylinder}).
The  perturbative result for the inside
of the cylinder is:
\begin{equation}
\frac{\Delta\gamma_{cyl}^{in}}{L}  =   \frac{1}{4} - \frac{1}{128}
\, \epsilon^2 + \cdots
\label{cyl}
\end{equation}

\begin{figure} [t]
\epsfxsize=4.0in
\epsfysize=4.0in
\centerline{\epsfbox{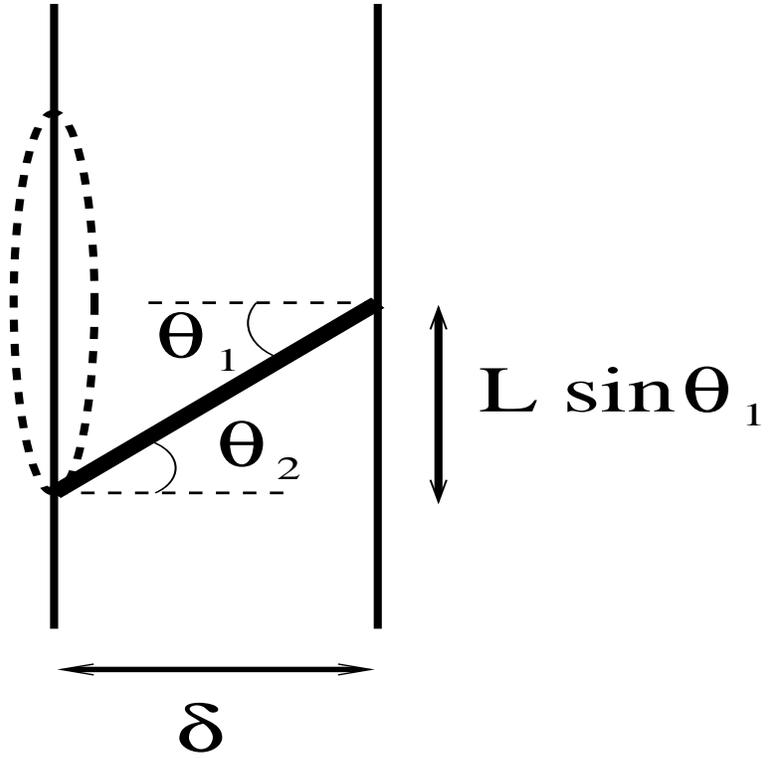}}
\vfil
\caption{Rods in a flat gap \--- side view.}
\label{figgap}
\end{figure}

The case of the flat gap has a different topology, and
therefore is worth doing as an example,
(see figure~(\ref{figgap}).)
For $L \leq \delta$, clearly there are no configurations of a rod that
bridge two points on the surface, and
therefore: $\Delta\gamma_{gap} = L/2$, (to assure consistency with
sections to come we took $S$ to be {\em one} side's area.) Whereas
for $L > \delta$, we have: $\theta_1=\theta_2=\theta$, $\cos\theta = \delta/L$,
and $\ell = 2 \pi L\, \sin\theta$. Thus:
\begin{equation}
{\partial^2\, \over \partial L^2} \Delta\gamma_{gap}=
-\frac{\cos^2\theta}{L} = - \frac{\delta^2}{L^3}
\label{secondderivativegap}
\end{equation}
It follows that:
\begin{equation}
\frac{\Delta\gamma_{gap}}{L} = \left\{
\begin{array}{lcl}
\delta/L - \delta^2/(2 \, L^2) & ; & L \geq \delta\\
1/2 & ; & L \leq \delta
\end{array}
\right.
\label{gap}
\end{equation}
This is to be compared to Eq.~(\ref{strip}) below.

We also developed a method to compute the small $L$ expansion of
the surface tension for a general surface in two and three
dimensions. We leave this discussion to appendix~\ref{app:smallL},
and just note here that the expansions of the expressions for the
circle, sphere, and the cylinder agree with this general method's
predictions. In particular, that $\Delta \gamma$ is an odd function
of $L$ for sufficiently small $L$ follows from this expansion.

\subsection{Random Insertion Sampling}
\label{sec:sampling}

We also tested the surprising prediction of equation(\ref{delgamcon3}) in
the case of an ellipsoid, the  simplest non-trivial case in 3 dimensions, by
performing random insertion sampling. Since the prediction for the surface
tension is in the limit of a non-interaction gas of rods, we need only consider
the behavior of one rod of length $L$ in our numerical experiment.

The algorithm is as follows.
We choose a volume $V_0$ around the ellipsoid of semi-axis $A,B,C$ 
that is large
enough to encompass the ellipsoid and a shell of thickness $L/2$ around the
ellipsoid. We then randomly select $M$ center of mass positions 
and orientations
for the rod of length $L$. The center of mass must be in $V_0$ excluding $V_e$,
the volume of the ellipsoid itself. By noting the fraction $f$ of insertions
that lead to overlap between the rod and the ellipsoid we can express the
surface-tension as:
\begin{equation}
  \Delta\gamma (L) = f \cdot \frac{V_0-V_e}{S_e}
\end{equation}
in units of $k_BT\rho_b$. $S_e$ denotes the surface area of the ellipsoid.

When we choose $M=10^8$, our numerical experiment confirms the prediction
$\Delta\gamma(L)/L = 1/4$ within an error-bar of $0.5*10^{-4}$ for a wide
variety of choices for $A, B, C,$ and $L$.
The error-bar on the fraction $f$ is $\sqrt{ \frac{f \cdot (1-f)}{M}}$

\subsection{Comments}
\label{sec:comments}

Although the method of ``bean-counting'' is less efficient than the
one described above, it is nevertheless useful for gaining more insight
into the reasons for the counterintuitive results above. In this method one
performs first the angular integration over $\rho({\vec{r}},{\vec{\omega}})$
to obtain the rod density $\rho({\vec{r}})$.
In the particular cases of flat, cylindrical and
spherical surfaces, the excess surface energy is then obtained from
\begin{equation}
\label{deltafcs} 
\Delta \gamma = \int d\,z \, (\rho_b-\rho(z)) \,J(z,R)
\end{equation} 
with $z$ the perpendicular distance from the surface, and $J(z,R)$
the Jacobian, which depends on the geometry. For flat surfaces
$J(z,R)=1$;  for the outside of a cylinder of radius R,
$J(z,R)= 1 + z/R$, and for a sphere $J(z,R)=(1 + z/R)^2$. For these
three geometries  there is only a one-dimensional
integral (over $z$)  to be carried out, once $\rho(z)$ is determined.
Differently from  the previous sections,
we define  hereafter the rod coordinates by the position of its center of mass.

\begin{figure} [t]
\epsfxsize=4.0in
\epsfysize=5.75in
\centerline{\epsfbox{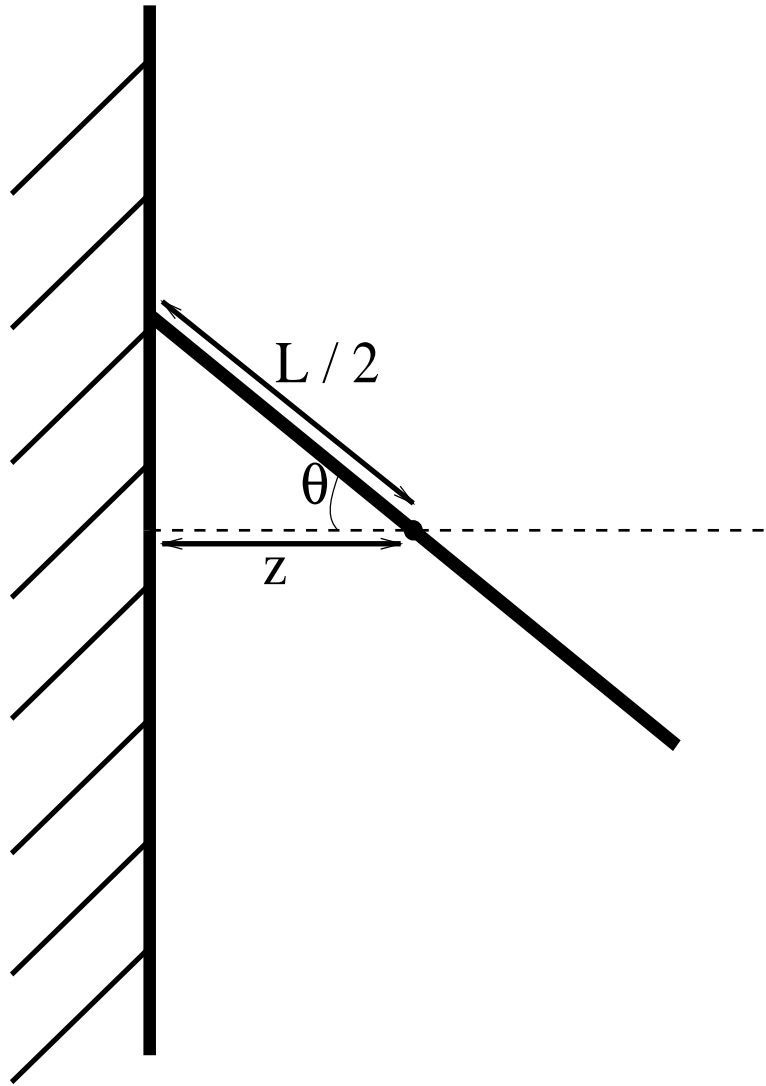}}
\vfil
\caption{Configurational space of  a rod near a flat surface,
The rod is represented at the angle of contact.
The complete allowed angular space is obtained by
rotation of the figure around the $z$-axis.}
\label{figflat}
\end{figure}

For example, for a plane with rods
on one side (see figure~(\ref{figflat})), we have:
\begin{equation}
\Delta \gamma_f= \int_0^{L/2} dz
 \left[1-{1\over{4\;\pi}}\int_0^{2\pi}d\phi
\int_{\arccos(2 z/L)}^{\pi-\arccos(2 z/L)}
   \sin\theta\;d\theta\;\right]  \;=\;\int_0^{L/2} dz \,
\left[1-\frac{2z}{L}\right]={L\over 4}
\label{planeone}
\end{equation}
Notice that the rod concentration $\rho=\rho_b\,  2 z/L$  is different from the
monomer concentration. The evaluation of this second quantity is slightly more
involved but can be calculated following Auvray
\cite{auvray}. This gives:
$c_f(z)\;=\;c_b \,(z/L) (1-\log(z/L))$,
where $c_b = \rho_b N$ is the bulk monomer concentration for a solution of
rods with $N$ monomers.

\begin{figure} [t]
\epsfxsize=4.8in
\epsfysize=3.8519in
\centerline{\epsfbox{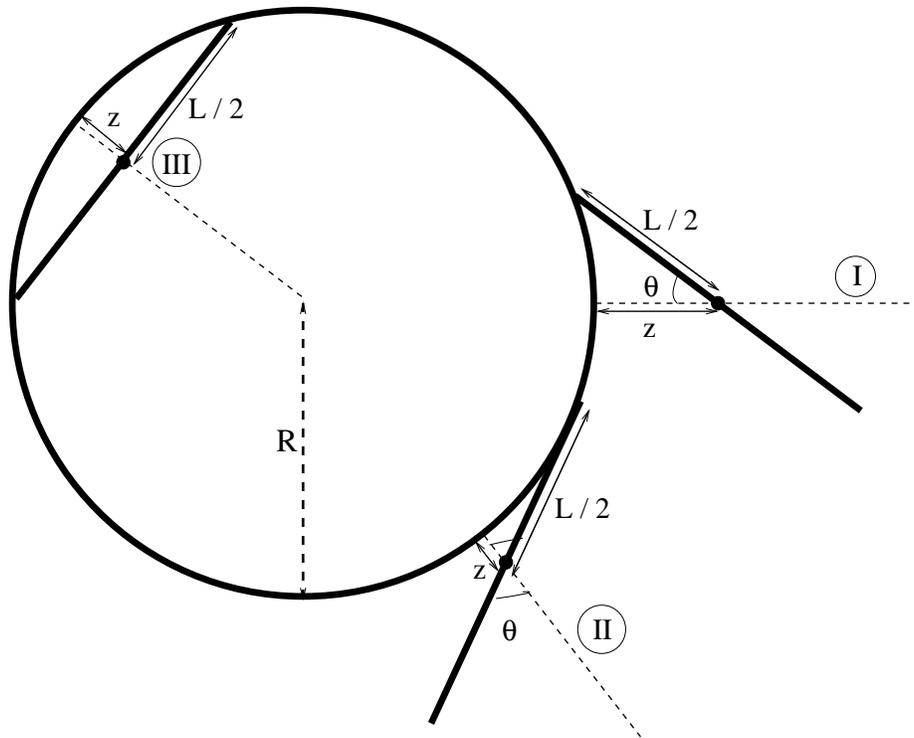}}
\vfil
\caption{Configurational space of rods in- and outside of a sphere.
I) $z > z_c^{out}$; II) $z < z_c^{out}$; III) $z = z_c^{in}$}
\label{figsph}
\end{figure}

The calculation of the phase space available to a rod in the presence
of a curved surface is somewhat different from the one
for a flat surface,  for now it
is possible for the rod to contact the wall in two different ways (see
figure~(\ref{figsph})): when
the rod is sufficiently far from the surface the tip touches
the wall upon rotation,
whereas when the rod is sufficiently close to the wall,
it will be tangent to
the surface at some point along its length. The critical
$z$ value separating these two regions is $z_{*,sph}^{out} =
\sqrt{(L/2)^2 + R^2} - R$.  The phase spaces arising from these two
configurations have different functional forms.  In the inside of the
sphere, of course, the tangent construction is  not possible, but now
there is a minimum distance beyond which it is not possible to place
the center of mass of the rod \--- this happens at
$z_{*,sph}^{in} = \sqrt{R^2 - (L/2)^2} - R$.  A closer look
at the results from the previous sections reveals  that the
surface tension is {\em non-analytic} in the curvature; e.g.
that even for arbitrarily small $\epsilon=L/R$, one
has neither $\Delta\gamma_{sph}^{out} (\epsilon)
= \Delta\gamma_{sph}^{in} (-\epsilon)$ nor $\Delta\gamma_{sph}^{out}
(\epsilon)
= -\Delta\gamma_{sph}^{in} (-\epsilon)$.
This invalidates the usual extrapolation of the free energy results from the
cylindrical  and the spherical shapes to arbitrary bending states. We can see
that the difference between $z_{*,sph}^{out}$, and $z_{*,sph}^{in}$
is the basic reason for this non-analyticity. It is also possible to
compute the  monomer densities in the spherical configuration,
we present these in appendix~\ref{app:sphere}.

\begin{figure} [t]
\epsfxsize=5.0in
\epsfysize=5.0in
\centerline{\epsfbox{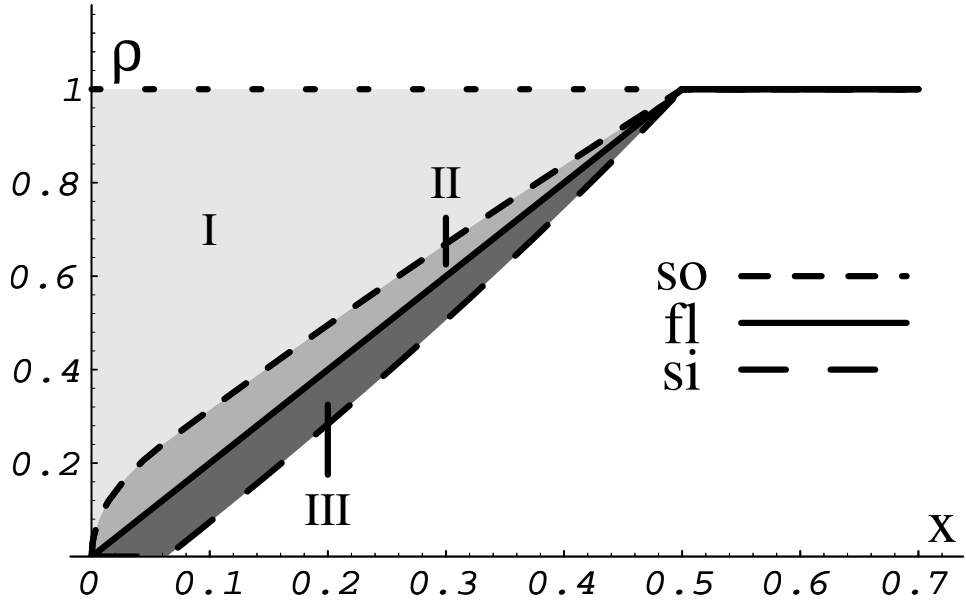}}
\vfil
\caption{The rod concentrations, $\rho(x)/\rho_b$,
compared for rods in- and outside of a sphere and
near a flat wall. The distances are in units of $L$.
{\bf so}: Outside of a sphere; {\bf fl}: Flat; {\bf si}: Inside a
sphere.
We took $R = 2\, L$ for this picture.}
\label{figden}
\end{figure}

\begin{figure} [t]
\epsfxsize=5.0in
\epsfysize=5.0in
\centerline{\epsfbox{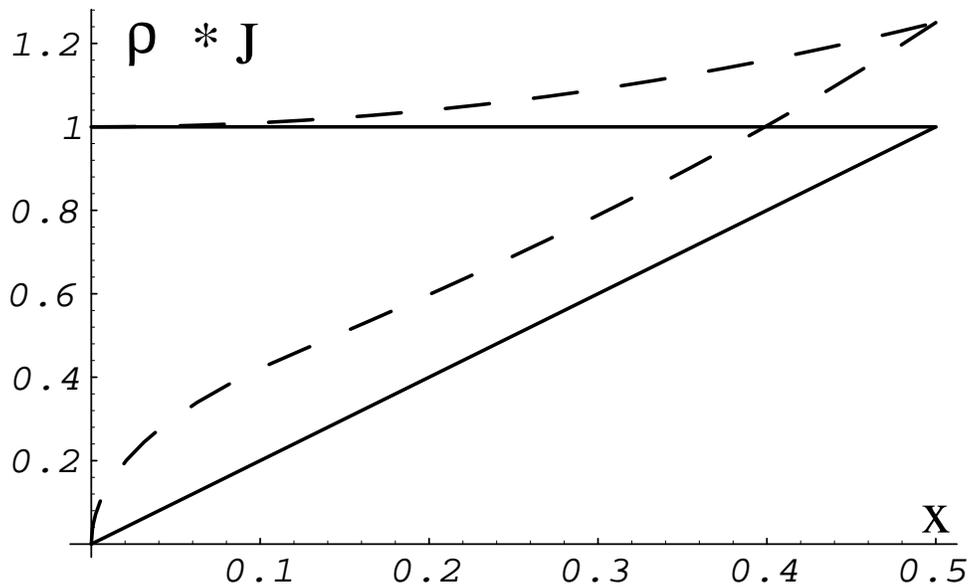}}
\vfil
\caption{This plot shows $\rho^{out}(x)/\rho_b$
weighted by the appropriate Jacobian, for the plane (solid line) and a
sphere of radius $R = L$ (dashed line).
The result in the text
corresponds to the equality of the area between the solid curves to
the one between the dashed curves.}
\label{figdenjac}
\end{figure}

From
equation~(\ref{planeone}) it is clear that differences between the excess
energy of a flat and a curved surface depend on two factors. The first
is the configurational part measured by the differences in the rod
density profiles, the second one is associated with the space
available to  the center of mass,
and is measured by $J(z,R)$. In the case of hard sphere solutions,
where there is no coupling between configuration and curvature, only
$J(z,R)$ is responsible for energy differences; the profile of
particle concentration, $\rho(z) = \rho_b \,
\Theta(z - r_0)$, is independent of geometry.
It is easy to show that
in that case one has $\Delta \gamma = k_B T \rho_b r_0 ( 1 + r_0/(2
R))$ for cylindrical surfaces and $\Delta \gamma = k_B T \rho_b r_0 (
1 + r_0/R + r_0^2/(3 R^2))$ for spherical ones.
In the case of rod solutions, there is a coupling between the configuration of
the rods and the surface curvature: the concentration profiles are a function
of the surface curvature. Given this coupling, the  exact
correspondence among  the free energies of all convex-shaped rod-systems,
proven in  section~\ref{sec:convex bodies}, is rather surprising. This
implies that curvature contributions to free energy(\ref{deltafcs})  from both
the profile and the Jacobian integrate out to a total vanishing value. This
cancellation  is non-local, i.e. at a given $z$  away from the surface the total
phase space for the center of mass  does not change as to compensate for the
configurational changes in the profile. One can gain further insight into the
issue by recalling equation~(\ref{deltafcs}), and considering
figure~(\ref{figden}). The surface tension is just the total ``area'' between
the curves $\rho(z)$ and
$\rho_b$ {\em weighted} by the appropriate Jacobian. The surface tension for
the plane is simply the sum of the shaded areas I and II, whereas the surface
tension for the outside of the sphere is the {\em weighted}  ``area'' of I
only, which happens to be the same.  The relevant weighted areas are shown in
figure~(\ref{figdenjac}).

Also, we pointed out that in general the surface tension decreases as
the surface develops concave regions, e.g.
$\Delta\gamma_{sph}^{in} < \Delta\gamma_f$,
i.e. the sum of the weighted
``areas'' of the regions I, II, and III in figure~(\ref{figden}) 
is {\em smaller} than the  area of I.
Therefore a flat surface with rods on one side would be able to lower
its energy by  bending {\em{towards}} the rods in a spherical shape;
and if there are rods on both sides, since $\Delta\gamma_{sph} <
\Delta\gamma_f$, they would destabilize the flat surface towards
bending.

\subsection{Bending Rigidities}
\label{sec:rigidites1}

For a flexible surface, e.g. a membrane, one computes the
renormalization of the bending rigidities by assuming that $\Delta f_{c}$,
the  excess free energy per unit area of the system is expandable
in a power series in the curvature, i.e. the excess free energy is analytic
in the curvature. Then up to quadratic order in the principal
curvatures ${1 \over {R_{1}}}$ and ${1 \over {R_{2}}}$,
this expansion can be written in
terms of the mean and Gaussian curvatures of the surface
\cite{safran}. In terms of
the principal radii, the mean curvature is \mbox{$H = {1 \over 2} ( {1
\over {R_{1}}} + {1 \over {R_{2}}})$}, and the Gaussian curvature is
\mbox{$K =  ({1 \over {R_{1}}})  ({1 \over {R_{2}}})$}.  A general
form of $\Delta f_{c}$ to this order is given by:
\mbox{$f_{c} = 2 \Delta \kappa ({H - c_{0}})^2 + \Delta {\overline{\kappa}}
K$},  where $\Delta \kappa$ and $\Delta {\overline{\kappa}}$ are the
contributions from the excess free energy to  $\kappa$ and
${\overline{\kappa}}$, the bare elastic constants of the membrane.  $c_0$ is
the induced  spontaneous curvature   of the system.

As an example we calculate first the modifications of  the elastic constants
of a membrane immersed in a dilute solution of hard spheres. As quoted above
$\Delta \gamma = k_B T \rho_b r_0 ( 1 + r_0/(2 R))$ for cylindrical surfaces
($1/R_1 = 1/R$ and $1/R_2 =0$) and $\Delta \gamma = k_B T \rho_b r_0 ( 1 +
r_0/R + r_0^2/(3 R^2))$ for spherical surfaces ($1/R_1 =1/R_2= 1/R)$.  This
leads to  $\Delta \kappa =0$, $\Delta \bar \kappa = (2/3) \,k_B T \, \rho_b
r_0^3$ and $c_0=0$.

For a membrane exposed to a rod solution, 
the non-analyticity of the free energy
might imply that the extracted  bending rigidities are not  meaningful for all
states of curvature of the membrane. In particular, bending states where the
principal radii of curvature have opposite signs might not be described by the
usual Helfrich expansion. With this note of caution in mind, we computed the
renormalization of the bending rigidities for a membrane exposed to a rod
solution on both sides:

\begin{equation}
\begin{array}{lclcl}
c_0 & = & 0\\
\Delta \kappa & = & - \frac{1}{64} \, \left(\rho_b \, L^3\right) & = &
		\frac{\rho_b}{\rho_b^*} \,
		\frac{1}{15.2} \, \frac{L}{D} \\
\Delta {\overline{ \kappa}} & =
	& + \frac{1}{96} \, \left(\rho_b \, L^3\right) & = &
		\frac{\rho_b}{\rho_b^*} \,
		\frac{1}{22.9} \, \frac{L}{D} \\

\end{array}
\label{kap}
\end{equation}

It is interesting that there is no spontaneous curvature
induced by the rods, since there is no symmetry reason why this should
be the case. Though, our ``small $L$ analysis'' in
appendix~\ref{app:smallL} shows that this is an (almost) general
property of rod-systems.
The modifications of the elastic constants (a decrease of $\kappa$ and an
increase of ${\overline{ \kappa}}$  indicating a preference to form  periodic
minimal surfaces) are similar to results for depletion of other macromolecular
species. However, the magnitude of this effect
can be here much larger than $k_B T$, a feature not found in other
depletion systems. It is of order of
$(\rho_b \, L^3) \simeq (\rho_b /\rho_b^*)\, (L/D)$,
where $D$ is the thickness of the rod and $\rho_b^*$
is the Onsager concentration for rigid rods.
Even at the Onsager concentration the enhancement factor $L/D$ can
lead to very large contributions.

\section{Two Surfaces in a Rod Solution}
\label{sec:twosurface}

Here we consider rods in shells where both surfaces of the shells are
repulsive. The shells are taken to be in equilibrium with a bulk reservoir
of a rod solution at bulk concentration $\rho_b$. The thickness of the shells
is denoted by $\delta$, the perpendicular distance from the inner surface by
$z$. The volume, and the thickness of the box are kept fixed as it gets bent.
When $\delta > L$, the results in this section follow from the ones in the
previous sections, since the effects of the two surfaces do not interfere with
one another. When $\delta < L$ however, there is interference and one needs to
calculate the allowed phase space by accounting for interactions with both
surfaces.

We consider rods confined inside flat or curved shells, but in
equilibrium with a rod solution.
By comparing the energies of flat, spherical and cylindrical shells,
we hope to investigate for rods a question that has recently been
investigated for flexible polymers\cite{yaman1} and mesoscopic quantum
systems\cite{jensen}: do the curved regions in confined systems act as
attractive potentials, where the concentration of particles will be greater, or
as depletive potentials from where particles will be repelled ?
Table~\ref{potentials} summarizes the systems studied so far,
including the case of confined rods treated below.

\subsection{Flat Strip}
\label{sec:flatstrip}

A straightforward calculation for a flat strip of thickness $\delta =
y \, L$ confining a rods solution yields:

\begin{equation}
\frac{\Delta\gamma^{in}_f}{L} = \left\{
\begin{array}{lcl}
y - \frac{1}{2} \, y^2 & ; & y < 1\\
\frac{1}{2} & ; & y > 1
\end{array}
\right.
\label{strip}
\end{equation}

\begin{figure} [t]
\epsfxsize=4.0in
\epsfysize=4.0in
\centerline{\epsfbox{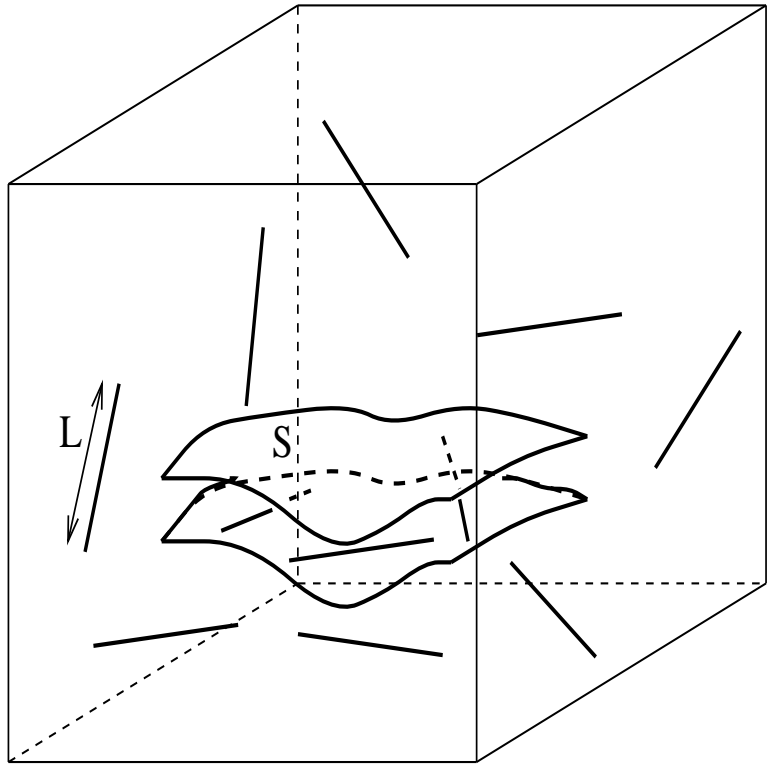}}
\vfil
\caption{Rods in a container with an open shell.}
\label{figbox}
\end{figure}

As usual, the interfacial energy is expressed in units of $k_B \, T \,
\rho_b $. Note that for separations larger than the rod-length the
interfacial energy is just the sum of two contributions from the
inside surfaces. When the surfaces are taken to be also in contact
with a solution outside (see figure~(.\ref{figbox})) we get:

\begin{equation}
\frac{\Delta\gamma^{both}_f}{L} = \left\{
\begin{array}{lcl}
\frac{1}{2} + y - \frac{1}{2} \, y^2 & ; & y < 1\\
1 & ; & y > 1
\end{array}
\right.
\end{equation}

\subsection{Spherical Shell}
\label{sec:sphshell}

We are able to calculate the available phase space, and thus the
surface tension of a spherical shell
exactly.  The result has four regions, across the boundaries of which
the functional form of $\Delta\gamma$ changes. This happens because one
needs to consider the relative positions of
$z_{*,s}^{out}$, $z_{*,s}^{in}$, which were defined
earlier,  and $z_0$, the point separating the two regions where the
rod only touches the upper surface and the one where it only touches
the lower one.
The extent of the first two
regions vanishes with curvature; the
results below tend asymptotically to the ones for the flat strip
in the limit of small curvature.
We took $S$ as the surface area of the flat box with the same volume
and thickness. We leave the full
results to the appendix and give here
the expansions in $\epsilon = L/R$.
\medskip
\begin{equation}
\Delta\gamma^{in}_{sph. sh.} = \left\{
\begin{array}{lcl}
y & ;   & y < \sqrt{\frac{1}{4} + \frac{1}{\epsilon^2}} -
\frac{1}{\epsilon}\\
y \left( 1 - \frac{2\,\sqrt{2}}{3} \,\sqrt{{y\, \epsilon}} +
   \frac{1}{2} \, \epsilon +
{1\over{3\,\sqrt{2}}} \,{{(y\, \epsilon)}^{{3\over 2}}} -\right.
& &\\
\hspace{0.75in}\left.{\epsilon^2} (\frac{1}{4} \, y - {{1}\over {48\,y}})
      + { {\rm O} (\epsilon) }^{5\over 2}\right)
& ; & \sqrt{\frac{1}{4} + \frac{1}{\epsilon^2}}
- \frac{1}{\epsilon} < y < \sqrt{1 + \frac{1}{\epsilon^2}} -
\frac{1}{\epsilon}\\
y - \frac{1}{2} \, y^2
+ \epsilon^2 (\frac{1}{24} \, y^4 + \frac{1}{48})
+ {\rm O} {(\epsilon)}^3
& ; &\sqrt{1 + \frac{1}{\epsilon^2}} - \frac{1}{\epsilon} < y < 1\\
\frac{1}{2}
+ \epsilon^2 (\frac{1}{12} \, y^2 - \frac{1}{48})
+ {\rm O} {(\epsilon)}^3
& ; & y > 1
\end{array}
\right.
\label{sphshellexp}
\end{equation}
\medskip

One can check by comparing Eq.(\ref{strip}) with
Eq.(\ref{sphshellexp}) that for any separation the surface tension of
the spherical shell is larger than that of the flat box.
We conclude therefore
that rods prefer to stay in flat regions of the confining
box rather than in spherically curved ones.
This can be further checked by computing the average rod concentration
inside the flat shell and comparing it to the value inside the
spherically bent shell; the latter will be smaller than the former.
We show below that
cylindrical shells induce the opposite, i.e. attract rods, for
$L$ less than but close to $\delta$.
Results for a spherical shell immersed in a a rod solution can be
easily obtained by adding to Eq.(\ref{sphshellexp}) the outside and
inside contributions of Eq.(\ref{sphere}).

\subsection{Cylindrical Shell}
\label{sec:cylshell}

The cylindrical shell calculation is substantially more
involved than the ones for planar or spherical shells. We resorted
to a perturbation in $(1/R)$ in this section, and stayed at
separations close to the rod-length, i.e. the results below for
$\delta < L$ should be understood to be valid only in a finite region
close to L.
Again, choosing $S$ as the flat box's surface area, we find:

\begin{equation}
\Delta\gamma^{in}_{cyl. sh.} = \left\{
\begin{array}{lcl}
1 - \frac{1}{2} \, y^2
+ \frac{1}{128} \,  \left(1 - 4 \, y^2 + 2 \,y^4 \right) \epsilon^2
+  \cdots & ; & y < 1\\
\frac{1}{2}
- \frac{1}{128} \, \epsilon^2 + \cdots & ; & y > 1
\end{array}
\right.
\label{cylshell}
\end{equation}

The results for the cylindrical shells are, at least for separations
close to the rod-length, opposite of those for the spherical shells:
the surface energy in this case is smaller than that for the flat
strip. Rod-like molecules will therefore be attracted by cylindrically
bent shells.

\subsection{Bending Rigidities}
\label{sec:rigidities2}

As explained above, for small curvatures, it is possible to describe
the elastic properties of a surface, by three parameters: the
spontaneous curvature $c_0$, and the bending rigidities $\kappa$, and
$\bar \kappa$.
In this paragraph we compute these parameters for a flexible box,
such as a membrane, exposed to rods. For all cases $c_0 = 0$, as it
should be due to the symmetry of the geometry.

For an open shell with rods on the inside only we find:

\begin{equation}
\begin{array}{l}
\Delta \kappa = - \frac{\rho_b}{\rho_b^*} \, \frac{L}{D} \, \left\{
\begin{array}{lcl}
\frac{1}{15.2} \, \left(- 1 \, + 4 y^2 - 2 y^4 \right) & ; & y < 1\\
\frac{1}{15.2} \,& ; & y > 1
\end{array}
\right.\\
\Delta \bar \kappa = + \frac{\rho_b}{\rho_b^*} \, \frac{L}{D} \,
\left\{
\begin{array}{lcl}
\frac{1}{22.9} \left(- 1 + 12 y^2 - 2 y^4 \right) & ; & y < 1\\
\frac{1}{22.9} \left(1 +  8 y^2\right) & ; & y > 1
\end{array}
\right.
\end{array}
\label{openin}
\end{equation}

For a strip immersed in a solution of rods, i.e. a
system with rods inside and outside the shell we find:

\begin{equation}
\begin{array}{l}
\Delta \kappa = - \frac{\rho_b}{\rho_b^*} \, \frac{L}{D} \, \, \left\{
\begin{array}{lcl}
\frac{1}{15.2} \, \left(- 1 + 4 y^2  - 2 y^4 \right)& ; & y < 1\\
\frac{1}{15.2} & ; & y > 1
\end{array}
\right.\\
\Delta \bar \kappa = + \frac{\rho_b}{\rho_b^*} \, \frac{L}{D} \,
\left\{
\begin{array}{lcl}
\frac{1}{22.9} \left(- 1 + 20 y^2 - 2 y^4 \right) & ; & y < 1\\
\frac{1}{22.9} \left(1 + 16 y^2 \right) & ; & y > 1
\end{array}
\right.
\end{array}
\label{openinout}
\end{equation}

\section{From Rigid Rods To Flexible Polymers}
\label{sec:segmented}

\begin{figure} [t]
\epsfxsize=4.667in
\epsfysize=4.0in
\centerline{\epsfbox{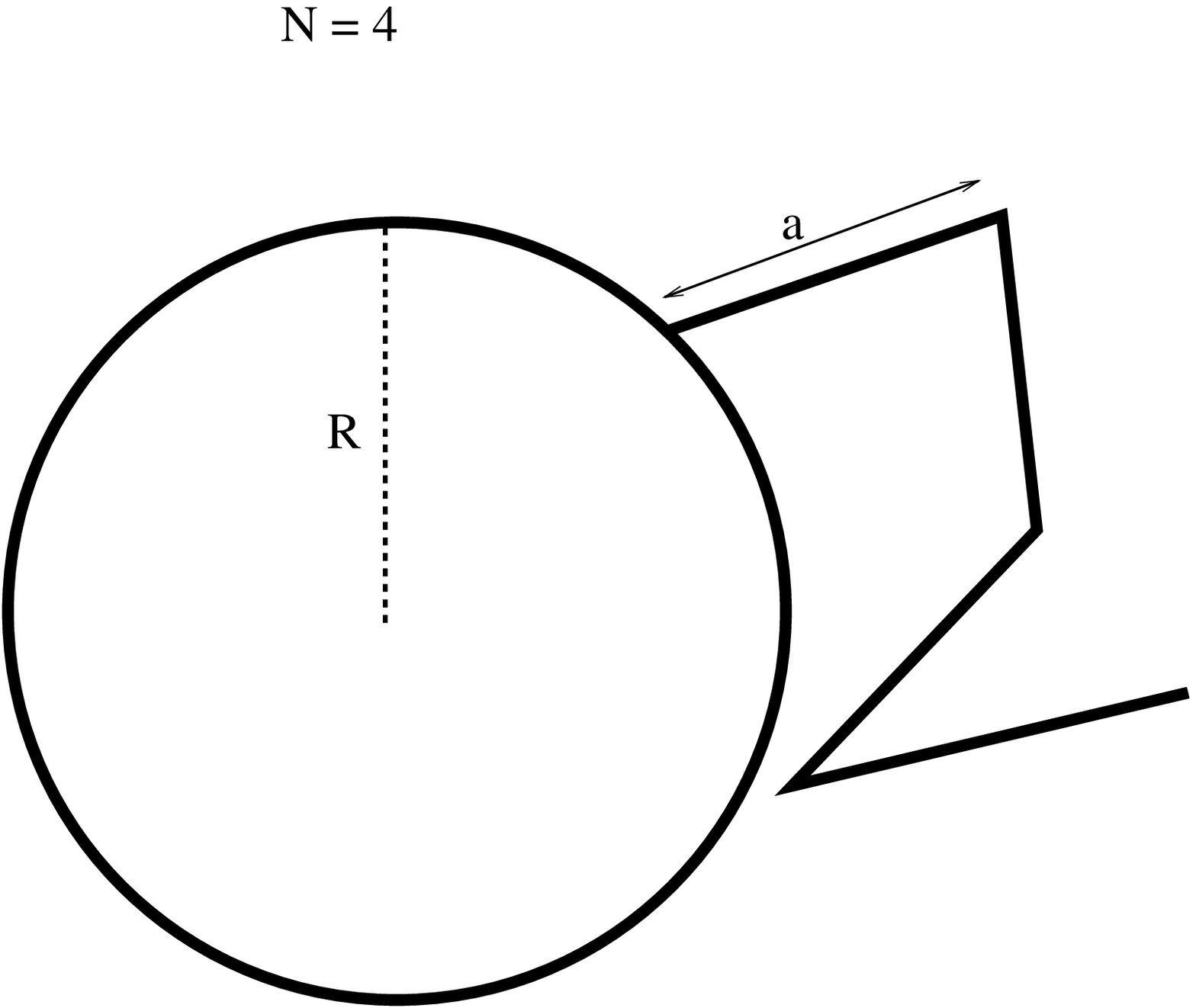}}
\vfil
\caption{Segmented rods near surfaces.}
\label{fighinged}
\end{figure}

A rigid rod can be viewed as the particular limit of a  freely-hinged
polymer chain with one monomer only, see figure~(\ref{fighinged}). 
In the reverse
limit, when the polymerization number $N$ is very large, the equilibrium
configurations of such a chain are well described by Gaussian statistics. We
now describe to some extent the crossover between these two limits.

For a membrane in contact with a dilute solution of Gaussian polymers, it is
possible to calculate the excess surface energy\cite{eisenriegler}. For
instance, a  sphere in $\cal D$ dimensions immersed in a such a solution
increases its surface energy by
\begin{equation}
\Delta \gamma_{poly} (N) = \frac{2}{\sqrt{\pi}} \, N^{1/2}\, a
+ \frac{{\cal D} - 1}{2} \, \frac{N\, a^2}{R}
+ \frac{({\cal D} - 1) ({\cal D} -3)}{6\, \sqrt{\pi}}\,
	\frac{N^{3/2}\, a^3}{R^2} + \cdots
\label{eisen}
\end{equation}
with $a$ the monomer size (and total length $L = N a$). For a sphere in three
dimensions this reduces to:

\begin{equation}
\Delta \gamma_{sph, poly} (N) = \frac{2}{\sqrt{\pi}} \, N^{1/2}\, a
+ \frac{N\, a^2}{R} +\cdots
\label{eisen2}
\end{equation}
Note that for the Gaussian polymer case, there is an analytical expansion of
the free-energy in powers of the curvature, the case of a solution inside a
sphere can simply be obtained from the result above by a sign permutation $1/R
\to -1/R$. 
The scaling of the first term in equation (\ref{eisen2}) implies that
for a plane the excess surface energy decreases by a factor of $\sqrt{2}$ when
one increases by a factor $2$ the number of links in a chain of length $L$.

In the rod limit ($N=1$), one has in the flat case, $\Delta\gamma_f (N = 1) =
(1/4) L$.  The ($N = 2$)-calculation for the flat case is straightforward and
gives:
\begin{equation}
\Delta\gamma_f (N = 2) = \frac{5}{24} L
\end{equation}
where $L = 2 N a$ is still the total length of the polymer. Interestingly,
even for such a small monomer number, doubling the number of monomers reduces
the surface tension by
$1.67$, a factor not too different from  $\sqrt{2}$  
in the Gaussian (large $N$)
limit. It is also possible to write down a general integral expression for the
surface tension near a flat wall for any $N$ where all  the integrals are in
principle doable; we left this to the appendix, Eq.~(\ref{flatN}).

The surface tension for ($N = 2$)  in- and
outside of a sphere can be calculated using  the same methods as for the rods
($N = 1$).  We leave the exact outside result to the appendix,
Eq.~(\ref{n2sph}),  and expand it here in the curvature:
\begin{equation}
\Delta\gamma^{out}_{sph} (N = 2)=
\frac{5}{24}
+ \frac{1}{64} \, \epsilon
- \frac{1}{240} \, \epsilon^2 + \cdots
\end{equation}

The exact inside result is simply:
\begin{equation}
\Delta\gamma^{in}_{sph} (N = 2) =
\frac{5}{24} - \frac{1}{64}\,\epsilon
-\frac{13}{1920}\, \epsilon^2
\end{equation}
Notice that  there is now a spontaneous curvature. Moreover, the
sign of it is positive for the outside, as it is for ideal polymers
\cite{eisenriegler}. Also, the asymmetry between the in- and the
outside cases, which is the cause of the non-analyticity in the curvature
free energy, is exactly one-third of what it is for ($N = 1$).
Furthermore,
the overall magnitude of the $\epsilon^2$ term is about half of what it is
for the rigid rod. This term does not exist for the polymer.
Therefore, even the introduction of one single link in the rigid rod, 
brings the results closer to those of the flexible polymer case.

\section{Conclusion}
\label{sec:conclusions}

We studied the depletion interactions of curved surfaces exposed to
dilute rod solutions. We find that in general, the excess surface energy
caused by the depletion interactions depends on the exact shape of the surface.
For instance, the surface tension of a plane exposed to a dilute  solution of
rods of length $L$ increases by $\Delta \gamma_{flat} = k_B T \rho_b L/4$, with
$\rho_b$ the rod number concentration of the solution, whereas the equivalent
quantity for a spherical surface in contact with such a solution {\sl inside}
the sphere is 
$\Delta \gamma_{sph}^{in} = k_B T \rho_b L/4 
\times ( 1 - (1/12)\,  L^2/R^2)$, 
with $R$ the radius of curvature of the sphere.

Surprisingly, we found also that a
spherical surface in contact with a rod solution {\sl outside} of the sphere
has the {\sl same}  excess surface energy as the flat surface $\Delta
\gamma_{sph}^{out} = k_B T \rho_b L/4$. Moreover we proved that this equality
holds for any convex surface exposed to the solution. This implies that the
excess  energy of any   convex body immersed in a dilute rod solution only
depends on the total surface of the body and not on its actual shape (as long
as it is convex). A similar equality also holds in any dimension of space. For
instance in two dimensions, the line tension for a straight line is $\Delta
\gamma_{flat} = k_B T \sigma_b L/\pi$ with $\sigma_b$ the ``bulk'' surface rod
density, and has the same value for rods outside of a circle.

The first practical implication of our results is that a flexible membrane
immersed in a solution (thus exposed to the depletion rod layer on both sides)
will spontaneously break its curvature symmetry. Also, a membrane exposed to a
rod solution (on one side) will spontaneously bend towards the solution.
The experimental observation of these effects will depend on the magnitude of
the symmetry breaking field. This can be measured by comparing the
relative importance of induced modifications in the elastic constants of the
membrane. Typical values of the bare elastic constants are in the range $1-20
k_BT$. The induced shift  to these  bare constants are for most systems like
hard spheres or flexible polymers, only a fraction of $k_B T$. However for rod
solutions, at the Onsager concentration, the induced modification is a factor
$L/D$ larger than $k_B T$, and can potentially be very large.

When the rods are confined in curved shells, we showed  that regions with
cylindrical curvature attract the rods, whereas regions with spherical
curvature deplete the rods. This enlightens a recent discussion on confinement
effects in solutions of  flexible polymers\cite{yaman1} and in mesoscopic
quantum systems\cite{jensen}. Table~\ref{potentials} summarizes results in a
variety of systems.

The methods that we developed for this work can be applied to other geometries
and other situations. Work is under progress to exactly evaluate  rod-mediated
sphere-sphere interactions. Also of interest are the  effects of attractive
interactions on the bending constants of membranes exposed to rod solutions.

We remark also that the steric surface interactions are
identical for rigid rods and rigid disks near surfaces,
if one takes the rod's length and the disk's diameter to be the same; however
the disk system does not enjoy the above-mentioned $L/D$ enhancement. The shell
calculations also generalize to the disks if the radius of the disks
is sufficiently small.

\section{Acknowledgments}
\label{acknowledgments}

KY and PP acknowledge support from NSF grants DMR-9624091 and
MRL-DMR-9632716. CMM was supported by CNRS and NATO fellowships, and
by the Petroleum Research Fund ($\sharp$29306-AC7), administered by the ACS.

\newpage
\appendix

\section{Small L expansion}
\label{app:smallL}

\begin{figure} [t]
\epsfxsize=4.0in
\epsfysize=4.0in
\centerline{\epsfbox{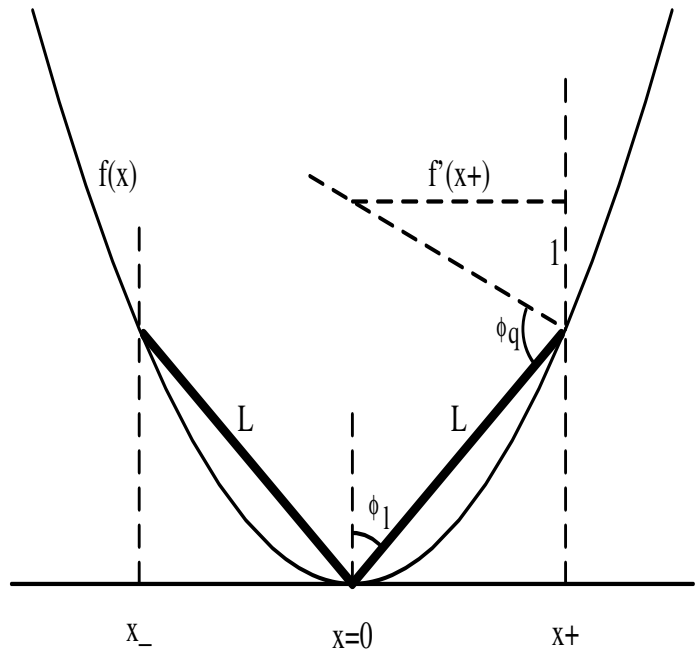}}
\vfil
\caption{This construction is used in the computation of the small $L$
expansion of the surface tension.}
\label{figtheorem2}
\end{figure}

When $L$ is small (or the radii of curvature large) we can expand the excess
free energy in powers of $L$.  We do this first for
the 2-D case. We start with Eq.~(\ref{ddelgamlrestwo}).
For $L$ small, $\sum_q\cos\phi_\ell\cot\phi_q$ is a function only of the
point $\ell$ and the expansion of the shape around it. We assume that $L$ is
small enough that rods of length $L$ starting at $\ell$ can only reach out to
points which are nearby along the surface. If the surface is convex in
this region, there are no points $q$, and the function is 0. We ignore
inflection points, because they will generally be of measure 0 (except
for plane regions, where the function is also 0). We assume that
there are no corners. This leaves only regions where the surface is
concave. Around the point $\ell$ we can draw the tangent to the surface as
the $x$-axis, with $\ell$ at $x=0$, and let the curve's shape be given by the
function $f(x)$, (See figure~(\ref{figtheorem2}).)
The function $f$ will have a Taylor expansion:
\begin{equation}
f(x)=c_2\,x^2 + c_3\,x^3 + c_4\,x^4+ \cdots
\label{smallL1}
\end{equation}
where the $c_i$ are functions of $\ell$.
We want to find the two values of $x$ such that rods of length $L$ start
at the origin and hit the curve (See figure~(\ref{figtheorem2}).)
That is, we need
solutions to $x^2+f(x)^2=L^2$. There are two solutions, which for $L$
small are near $\pm L$. We can expand in powers of $L$, and to
$4^{\rm{th}}$ order find:
\begin{equation}
x_{\pm}=\pm L \mp {1\over 2}c_2 \,L^3 - c_2\,c_3\,L^4 + O(L^5)
\label{smallL2}
\end{equation}
It is not hard to show
that the two solutions $x_{\pm}$ are given by the same power series,
except with the sign of the odd powers in $L$ reversed. Another way to say
this is that the two values of $x$ are given by one power series in $\pm
L$. Given the shape $f(x)$ and values $x_{\pm}$, we can calculate the cosines
and cotangents (see figure~(\ref{figtheorem2})):
\begin{equation}
\sum_q\cos\phi_\ell\cot\phi_q =
  \sum_{x=x_{\pm}} {f\over L}{{xf'-f}\over{L\sqrt{1+(f')^2}}}\Bigg{/}
  \sqrt{1-{{(xf'-f)^2}\over{L^2(1+(f')^2)}}}
\label{sum2d}
\end{equation}
Plugging in the expansion of $x_{\pm}$ to order $L^3$ we get:
\begin{equation}
\begin{array}{lcl}
\cos\phi_\ell\cot\phi_q & = & c_2^2 \,L^2 \pm 3\, c_3\, L^3 + O(L^4) \\
\sum_q\cos\phi_\ell\cot\phi_q & = & 2\, c_2^2 \,L^2 + O(L^4)
\end{array}
\label{smallL3}
\end{equation}
In the sum over $x_{\pm}$, the terms of order $L^3$ canceled out.
Because $x_{\pm}$ is a function of $\pm L$, functions of $x_{\pm}$ are
also functions of $\pm L$.
The expression for
$\cos\phi_\ell\cot\phi_q$ above uses only functions of $x_{\pm}$ and
$L^2=(\pm L)^2$,
so is again a function of $\pm L$. It follows that when we sum over $q$,
all terms odd in $L$ will cancel out in $\sum_q\cos\phi_\ell\cot\phi_q$,
so ${{{d^2\Delta\gamma}\over dL^2}}$ and $\Delta\gamma$ are odd
functions of $L$ for $L$ small!
If we put everything together, and integrate, we get:
\begin{equation}
\begin{array}{lcl}
\Delta\gamma(L) & = & \rm{an\;odd\;function\;of\;L} \\
 & = & {L\over\pi} - {{L^3}\over{24\pi}}\overline{({1\over{R^2}})} + O(L^5)
\end{array}
\label{delgam2dsmallL}
\end{equation}
Here $R$ is the local radius of curvature, and
$\overline{({1\over{R^2}})}$ indicates that we are averaging ${1\over{R^2}}$
over the perimeter of the object, weighting by arclength. We set
${1\over{R^2}}$ to 0 where the object is convex. The assumption that there
are no corners is necessary -- for example, for rods inside a rectangle there
will be terms of order $L^2$. And if $L$ is too large, we can no longer do this
small $L$ expansion and  $\Delta\gamma$ does not have
to be an odd function of $L$ -- for example, if we are inside a closed body,
and $L$ is larger than all lengths of the body,
$\Delta\gamma$ is simply a nonzero constant independent of $L$.

The expansion in 3 dimensions is a little more complicated, but
essentially the same. We again assume that there are no corners, and
that inflection points outside of plane regions are a set of measure 0.
We have a tangent $xy$-plane and a function $f(x,y)$.
We expand $f(x,y)$ in powers of $x$ and $y$, starting at $x^2$ and $y^2$
and rotating out the $xy$ term. The solutions to $x^2 + y^2 + f(x,y)^2 =
L^2$ are a near-circle.
Letting $\phi$ be the angle in the $xy$ plane, and putting
$u^2\equiv x^2+y^2$ we have
\begin{equation}
\begin{array}{rcl}
f(u,\phi) & = & c_{2,\phi}\,u^2 + c_{3,\phi}\, u^3
		+c_{4,\phi}\, u^4+\cdots \\
c_{2,\phi} & \equiv & c_{20}\cos^2\phi+c_{22}\sin^2\phi \\
c_{3,\phi} & \equiv & c_{30}\cos^3\phi+c_{31}\cos^2\phi\sin\phi+
  c_{32}\cos\phi\sin^2\phi+c_{33}\sin^3\phi
\end{array}
\label{smallL4}
\end{equation}
Note that $c_{n,\phi}=(-)^nc_{n,\phi+\pi}$. We now need to look for
solutions of $u^2+f(u,\phi)^2=L^2$ for each particular $\phi$. But this is
just the 2-D problem, which has already been done. The only additional
complication is the arclength factor, which can be rewritten in
spherical coordinates as:
\begin{equation}
\begin{array}{lcl}
d\ell & = & \sqrt{(L\;d\theta)^2+(L\sin\theta\;d\phi)^2} \\
 & = & d\phi \, L\, \sqrt{1-\cos^2\theta +
   {1\over\sin^2\theta}\left({d(\cos\theta)\over{d\phi}}\right)^2} \\
 & = & d\phi \, L\, \sqrt{1-{\left({f\over L}\right)}^2+
   {1\over{1-{({f\over L})}^2}}
   {\left({d\over{d\phi}}\left({f\over L}\right)\right)}^2}
\end{array}
\label{arclength}
\end{equation}
The values of $f$ at angles $\phi$ and
$\phi+\pi$ have the same power series in $\pm L$, and $L$ only appears
independently in $L^2 = (\pm L)^2$. So as in the 2-D case,
$\Delta\gamma$
is an odd function of $L$ for $L$ small.
The low order expansion in $L$ gives:
\begin{equation}
\begin{array}{lcl}
\int d\ell\cos\theta_1\cot\theta_2 & =&
   L\int_0^{2\pi}
	d\phi\, [c_{2\phi}^2 \,L^2 + c_{3\phi}\,L^3 + O(L^4)] \\
   & = & L^3{\pi\over 4}(3\, c_{20}^2 + 2\, c_{20}\, c_{22}
		+ 3\, c_{22}^2) + O(L^5)
\end{array}
\label{smallL5}
\end{equation}
Integrating twice and rewriting in terms of the principal
radii of curvature,
\begin{equation}
\Delta\gamma(L) = {L\over 4} - {L^3\over 384}\,
\overline{\left({3\over{R_1^2}}+
                        {2\over{R_1R_2}}+{3\over{R_2^2}}\right)} + O(L^5)
\label{delgam3dsmallL1}
\end{equation}
Once again, when taking this average it is understood that
we take the quantity inside the parentheses to be zero if the surface is
concave at the point at hand.
Now in terms of the Gaussian and the mean curvatures:
\begin{equation}
\begin{array}{lcl}
\Delta\gamma(L) & = & \rm{an\;odd\;function\;of\;L} \\
  & = & {L\over 4} - {\overline {\left(\frac{H^2}{32} -
 	\frac{K}{96}\right)}}\, L^3 + O(L^5)
\end{array}
\label{delgam3dsmallL2}
\end{equation}

\section{Sphere and Spherical Shell}
\label{app:sphere}

The probability of finding a rod's center of mass $z$ away from a
sphere can be calculated by first computing the allowed angular phase
in figure~(\ref{figsph}) and then integrating over the angles.
For convenience we define $r \equiv R/L = 1/\epsilon$, and still use
$x \equiv z/L$. We find for these probabilities:

\begin{equation}
\begin{array}{lcl}
\rho^{out}_{sph}(x)/\rho_b & = &
	\left\{ \begin{array}{lcl}
		\sqrt{2 r x + x^2} / (r + x) & ; & 0 < x < x_{*,sph}^{out}\\
		2 x + (1/4 - x^2)/(r + x) & ; & x_{*,sph}^{out} < x < 1/2 \\
		\end{array}
	\right.
\\

\rho^{in}_{sph}(x)/\rho_b & = &
	\left\{ \begin{array}{lcl}
		0	& ; & 0 < x < x_{*,sph}^{in}\\
		2 x - (1/4 - x^2)/(r - x) & ; & x_{*,sph}^{in} < x < 1/2 \\
		\end{array}
	\right.
\end{array}
\label{densph}
\end{equation}
		
The monomer concentration in- and out-side of a sphere can be computed
following the construction in \cite{auvray}.
We define
$\lambda_*(r, x) \equiv 1/2 - \sqrt{(r-x)^2 - r^2 + (1/2)^2}$:\\
For the inside of the sphere the concentration profile is given
by\footnote{Strictly speaking this calculation assumes $L <= R$.}:

\begin{equation}
c_{sph}^{in}(x) =  \left\{
\begin{array}{lccl}
c_{sph, I}^{in}(x) & &; & 0 < x < r - \sqrt{(1/2)^2 + r^2}\\
c_{sph, II}^{in}(x) & &; & r - \sqrt{(1/2)^2 + r^2} < x < 1\\
c_b & &; & x > 1
\end{array}
\right.
\label{consphin}
\end{equation}
where
\begin{equation}
\begin{array}{lcl}
c_{sph, I}^{in} (x)  & \equiv & c_b \, \left[x - x \log{(x)} -
{\frac{1}{2 (r - x)}} \left({\frac{1}{2}} (1 - x^2) + x^2 \log{(x)}
\right) + \right. \\
& & \left. x {\log{\left(\frac{\lambda_*}
{1 - \lambda_*}\right)}} - \frac{1}{2 (r - x)} \left(\frac{1}{2}
(\lambda_*^2 - (1 - \lambda_*)^2) - x^2 {\log{\left(\frac{\lambda_*}
{1 - \lambda_*}\right)}} \right) \right]\\
c_{sph, II}^{in} (x) & \equiv &
c_b \left[ x - x \log{(x)} - \frac{1}{2 (r - x)} \left(\frac{1}{2}
(1 - x^2) + x^2 \log{(x)} \right) \right]
\end{array}
\label{conspheinapp}
\end{equation}
The monomer concentration on the outside of a sphere
is\footnote{Strictly speaking this calculation assumes $L <= (4/3) R$}:

\begin{equation}
c_{sph}^{out}(x) =  \left\{
\begin{array}{lccl}
c_{sph, I}^{out} (x) & &;  & 0 < x < \sqrt{1 + r^2} - r\\
c_{sph, II}^{out} (x) & &; & \sqrt{1 + r^2} - r < x < 1\\
c_b & &; & x > 1
\end{array}
\right.
\label{consphout}
\end{equation}
where

\begin{equation}
\begin{array}{lcl}
c_{sph, I}^{out} (x) & \equiv &  c_b \left({1 \over {4\, \left( r + x
\right) }} \right) \left( -2\,r\,x + 4\,{\sqrt{x\,\left( 2\,r + x
\right) }} - (2 \,r \,x + x^2) \, \log{\left(1 + \frac{2 \, r}{x}
\right)} \right)\\
c_{sph, II}^{out} (x) & \equiv & c_b  \left({1
\over{4\, \left( r + x \right) }} \right) \left(1 + 4\,r\,x + 3\,{x^2}
- 2\, (2 \,r \,x + x^2)\, \log{(x)} \right)
\end{array}
\label{conspheoutapp}
\end{equation}

Notice that the first regimes in both cases disappear as the radius
becomes large, and one can readily check that the second regime's
profile goes to the flat case's profile in this limit.  This
concentration profile calculation, for the outside is done in
\cite{auvray} as well, where the author also finds these three
regimes. Our result agrees with his in regimes II and (trivially)
III. But we differ in regime I. Since his result is not continuous
across the boundary between I and II we believe that his
calculation for that case is inaccurate.
Our result is continuous, and
furthermore the full profile integrates to the full partition function
for this geometry, which is a
non-trivial consistency check. The profile for the inside is also
continuous and integrates to the partition function for the inside of
the sphere \--- again a non-trivial consistency check.

The full, exact expression for the surface tension in a spherical
shell is given by:
\begin{equation}
\Delta\gamma^{in}_{sph} =
y \, \left(1 - \frac{\tilde{v}(y, \epsilon)}{v(y, \epsilon)}\right)
\end{equation}
where
$v(y, \epsilon) \equiv y
+ \epsilon \, y^2
+ \frac{1}{3} \, \epsilon^2 \, y^3$,
and

\begin{equation}
\tilde{v}(y, \epsilon)= \left\{
\begin{array}{lcl}
0
& ; & y < \sqrt{\frac{1}{4} + \frac{1}{\epsilon^2}} -\frac{1}{\epsilon}\\
\frac{1}{16}
\left( 8 \, \epsilon \, y
- 4\,\epsilon\, {\sqrt{\epsilon\,y\,\left(2 + \epsilon\,y\right)}}
+ 4\, \epsilon^2 \, y^2  + \epsilon^2
\,\right) +
& &\\
\frac{1}{24}
{\left( 8 \, \epsilon \, y
- 4\,\epsilon\, {\sqrt{\epsilon\,y\,\left(2 + \epsilon\,y\right)}}
+ 4\, \epsilon^2 \, y^2  + \epsilon^2 \,\right)}^{3/2}
& ; &
\sqrt{\frac{1}{4} +
\frac{1}{\epsilon^2}} < y + \frac{1}{\epsilon} < \sqrt{1 +
\frac{1}{\epsilon^2}}\\
\frac{1}{2}\, y^2 + \frac{1}{8} \, \epsilon^2\,y^4
-  \frac{1}{48}\, \epsilon^2 + \frac{1}{2}\, \epsilon\, y^3
& ; & \sqrt{1 + \frac{1}{\epsilon^2}} - \frac{1}{\epsilon} < y < 1\\
y - \frac{1}{2} + \epsilon \, y^2 - \frac{1}{2} \,\epsilon \, y
+ \frac{1}{3} \epsilon^2 \, y^3 - \frac{1}{4} \epsilon^2 \, y^2
+ \frac{1}{48} \epsilon^2
& ; & y > 1
\end{array}
\right.
\label{sphshell}
\end{equation}

\section{Segmented Rod Results}
\label{app:segmented}

The integral expression for the surface tension of an $N$-polymer
around a flat surface is:

\begin{equation}
\Delta\gamma_f =
N - (1/2)^N \, \int_0^{N} \, dz_1
\prod_{i = 1}^{N} \,
\left(\int_{-1}^{1} \, du_i \left[
\Theta (z_i - 1) + \Theta (1 - z_i) \, \Theta (z_i - u_i) \right]
\right)
\label{flatN}
\end{equation}
where $z_i \equiv z_1  - {\displaystyle\sum_{j = 1}^{i - 1}} u_j$, and
$\Theta$ is the step-function.
This formula gives the same answers as our intuitive way of doing
these integrals for ($N = 1$), and ($N = 2$.)
With sufficient patience or
efficient programming it should be straightforward to see how the
result changes as $N$ grows larger.

The exact expression for the surface tension for ($N = 2$)
outside of a sphere is:

\begin{equation}
\Delta\gamma^{out}_{sph} =
\frac{2}{15} \, \left(\frac{1}{\epsilon}\right)^3 \,
\left[ \left(1 + \frac{1}{4}\,\epsilon^2\right)^{5/2} -
\left(1 + \frac{5}{8}\, \epsilon^2 - \frac{25}{16}\,\epsilon^3
+\frac{1}{32}\, \epsilon^5\right) \right]
\label{n2sph}
\end{equation}

\newpage

\newpage

\newpage

\newpage

\newpage

\newpage

\newpage

\newpage

\newpage

\newpage

\newpage

\newpage

\begin{table} [t]
\begin{tabular}{lll}
System & Spherical & Cylindrical\\
\hline
Quantum Particle & Neutral & Attraction\\
Gaussian Chain & Neutral & Attraction\\
Hard Spheres & Repulsion & Neutral\\
Rods & Repulsion & Attraction
\end{tabular}
\vspace{0.2cm}
\caption{Table of potentials induced by curvature in several systems.}
\label{potentials}
\end{table}

\end{document}